\documentclass[journal=jctcce,manuscript=article]{achemso}

\usepackage[version=4]{mhchem}

\usepackage{tikz}
\usetikzlibrary{quantikz2}

\author{Alexandre Fleury}
\email{alexandre.fleury@sandboxaq.com}
\affiliation[Sandbox]{SandboxAQ, Palo Alto, CA 94301, USA}
\author{James Brown}
\affiliation[qBraid]{qBraid Co., Chicago, IL 60615, USA}
\altaffiliation{The work of J.B. was performed before joining qBraid.}
\author{Erika Lloyd}
\author{Maritza Hernandez}
\affiliation[Sandbox]{SandboxAQ, Palo Alto, CA 94301, USA}
\author{Isaac H. Kim}
\affiliation[UDavis]{Department of Computer Science, University of California, Davis, CA 95616, USA}

\title{Non-unitary Coupled Cluster Enabled by Mid-circuit Measurements on Quantum Computers}

\begin{document}

\begin{abstract}
  Many quantum algorithms rely on a quality initial state for optimal performance. Preparing an initial state for specific applications can considerably reduce the cost of probabilistic algorithms such as the well studied quantum phase estimation (QPE). Fortunately, in the application space of quantum chemistry, generating approximate wave functions for molecular systems is well studied, and quantum computing algorithms stand to benefit from importing these classical methods directly into a quantum circuit. In this work, we propose a state preparation method based on coupled cluster (CC) theory, which is a pillar of quantum chemistry on classical computers, by incorporating mid-circuit measurements into the circuit construction. Currently, the most well studied state preparation method for quantum chemistry on quantum computers is the variational quantum eigensolver (VQE) with a unitary-CC with single- and double-electron excitation terms (UCCSD) ansatz whose operations are limited to unitary gates. We verify the accuracy of our state preparation protocol using mid-circuit measurements by performing energy evaluation and state overlap computation for a set of small chemical systems. We further demonstrate that our approach leads to a reduction of the classical computation overhead, and the number of CNOT and T gates by 28\% and 57\% on average when compared against the standard VQE-UCCSD protocol.
\end{abstract}

\section{Introduction}

Many of the major challenges we face today could be solved or mitigated with the advent of
new specialized functional materials and drugs. This requires tremendous research effort to design, explore and examine potential molecular candidates, as the number of possible compounds is intractable for a brute-force search~\cite{finkVirtual2005}. Simulating chemistry \emph{in silico} may bring considerable speed-up in the development process by screening out molecules with undesirable properties without having to synthesize them in laboratory. To model the quantum mechanical behavior of molecules, one must work with the Schrödinger's equation, a linear partial differential equation. In the majority of instances, the time-independent version of it, represented by

\begin{equation}\label{eq:timeind_schro}
  \hat{H}\ket{\psi} = E \ket{\psi},
\end{equation} 

is sufficient to observe correlation between steady-state molecular properties to simulated quantities. Here, $\hat{H}$, $\ket{\psi}$ and $E$ refer to the Hamiltonian, the wave function and the eigenvalue of $\hat{H}$, respectively. Usually, the Hamiltonian incorporates the Born-Oppenheimer approximation~\cite{born1985quantentheorie}, which treats the nuclear and electronic degrees of freedom separately, such that the $E$ refers to the total electronic energy. As stated in the quantum postulates, the wave function $\ket{\psi}$ encodes all the molecular characteristics: a particular molecular property can be extracted using the appropriate quantum operator.

An exponential computational complexity emerges with classical algorithms when scaling up the problem complexity, i.e. increasing the number of atoms, considering heavier atoms, or using more accurate algorithms. In fact, the most elaborate molecules to be exactly modelled within the boundaries of the Born-Oppenheimer approximation have been limited to few-atom species~\cite{rossiFullconfiguration1999,ganLowest2006}. Industry-relevant catalysts or polypeptides are way beyond the capabilities of the most powerful conventional computing devices. This is true even for the widely used coupled cluster (CC) method in classical quantum chemistry (equations are presented in Section~\ref{sec:cc}). In fact, the CC with a full treatment of singles and doubles, with estimation of connected triples contribution using many-body perturbation theory arguments - CCSD(T) - is the gold standard in the community, and scales as $\mathcal{O}(N^7)$, where $N$ is the number of basis functions.

Simulating quantum effects on quantum computers is anticipated to be an application where quantum utility could be observed. Mimicking quantum behavior, especially the real time evolution of molecular wave functions (dynamics), scales more efficiently on quantum devices as demonstrated in studies related to Hamiltonian simulation~\cite{babbushQuantum2023,childsHamiltonian2012,LoaizaRecuding2023,LowOptimal2017}. To confirm a practical quantum advantage, one must reveal the prefactors not taken into account in the big O notation, which is a non-trivial task. Moreover, recent results suggest the absence of a quantum advantage for quantum chemistry ground-state calculation~\cite{liuCan2023,leeEvaluating2023}, and the 2-local Hamiltonian problem has been shown to belong to the Quantum Merlin Arthur (QMA) complete class~\cite{kempe2006complexity}.

Quantum phase estimation (QPE) is a quantum algorithm capable of computing the eigenvalue of an arbitrary state described by a given Hamiltonian. The cost of this algorithm depends on three factors: (1) the targeted precision that depends on the number of ancilla qubits, (2) the gate cost for implementing the Hamiltonian, and (3) the overlap $\epsilon$ between the input state and the targeted state~\cite{leeEvaluating2023}. While the first factor is trivial, numerous studies have focused on reducing the cost of Hamiltonian simulation~\cite{LowOptimal2017,berry2015simulating,low2019hamiltonian,liuCan2023}, achieving near-optimal scaling depending on the Hamiltonian structure. The remaining factor is commonly referred to as state preparation. For simple molecular systems, the Hartree-Fock state $|\psi_0\rangle$ significantly overlaps with the true ground state, but this is not the case for complex molecular systems~\cite{kohn1999nobel,chan2012low}. As demonstrated by \citeauthor{PhysRevA.107.L040601}, the overhead cost for preparing a better initial state is worth the effort, as QPE is an expensive quantum oracle~\cite{PhysRevA.107.L040601}.

State preparation is an active research field in quantum computing that aims to construct circuits efficiently, with significant overlap with target states. One well studied approach for initial state preparation for chemistry applications is the Variational Quantum Eigensolver (VQE) with a Unitary Coupled Cluster (UCC) ansatz~\cite{peruzzo2014variational,anandQuantum2022}. To obtain the minimal eigenvalue of the Hamiltonian for a given ansatz, classical optimization must be performed. As a result, VQE encounters multidimensional optimization difficulties~\cite{mcclean2018barren}. An additional challenge with the VQE approach is that the eigenvalue (the molecular energy) must be sampled often during the optimization process. Considering that even a single energy evaluation is a computationally expensive task, VQE runtimes are expected to become intractable with the system size~\cite{gonthierMeasurements2022}.

Many UCC-inspired ansatzes have been published~\cite{peruzzo2014variational,PhysRevA.103.032605,PhysRevA.98.022322,PhysRevA.95.020501}, and they can be easily mapped into a quantum circuit composed of unitary gates. These unitary gates are the only permitted operations on a quantum device because of the time evolution postulate. The exception to this statement is measurements, which cause the collapse of the wave function. Recent studies show that long-range entanglement can be generated on a set of qubits more efficiently using measurement gates than with pure unitary operations~\cite{baumerEfficient2023,foss-feigExperimental2023}. This sparked the idea of mapping non-unitary coupled clusters equations, which are commonly implemented in quantum chemistry packages, on quantum computers. This non-unitary concept has been considered by modifying the operators, as demonstrated by the method of moments of coupled cluster equations (MMCCs)~\cite{peng2022mapping}. \citeauthor{erhart2024chebyshev} also reported recently employing Chebyshev polynomials to map variational coupled cluster (VCC) states on quantum devices using quantum singular value transformation (QSVT)~\cite{erhart2024chebyshev}. Our approach is distinct as it is can be roughly summarized as an implementation of a trotterized CC, and each cluster operator is implemented as Linear Combination of Unitaries (LCU)~\cite{childsHamiltonian2012} primitives on an identity operator and a fermionic excitation operator. This way of mapping the classical CC amplitudes circumvents the steps of repeated energy evaluations, at the cost of having a probabilistic quantum state preparation. Those results aim at providing insights into the research question of "How can we make use of measurement gates to create efficient entanglement between qubits for chemistry problems?"

The idea of initializing arbitrary classical data on a quantum device is commonly referred to as QRAM, and is a generalization of state preparation in a wider range of applications in quantum computing~\cite{tubmanPostponing2018,deverasDouble2022,zhangQuantum2022}. Another study also shows a strategy of block encoding to encode sum of Slater determinants (SOS) and matrix product state (MPS) for chemistry applications~\cite{fomichevInitial2023}. The main takeaway is that the most appropriate state preparation protocol is likely to depend on the specific use case. In fact, each protocol has its own cost in classical preprocessing, CNOT count, circuit depth, and ancilla qubits requirement.

We propose a workflow that exposes the utility of mid-circuit measurements in the implementation of coupled cluster states on quantum computers, while circumventing the drawbacks of variational optimization. The research article is divided as follows. In Section~\ref{sec:cc}, we present the construction of the quantum circuit primitives for the coupled cluster equations. We then demonstrate how the low-level primitives are used to construct the full quantum circuit. In Section~\ref{sec:results}, we reveal results about mapping pre-computed coupled cluster amplitudes onto quantum circuits, and a discussion about classical and quantum scaling of this workflow. The article ends with a scaling analysis of the classical computation cost, and the quantum circuit complexity.

\section{From coupled cluster equations to quantum circuits}\label{sec:cc}

In this section, we show the thought process behind the circuit primitives construction. A summary of the coupled cluster equations is presented, and the simplest cluster operator mapped to a quantum circuit is explained. Finally, the full circuit construction using mid-circuit measurements is shown.

\subsection{Coupled cluster}

There are multiple approaches for solving Eq.~\ref{eq:timeind_schro}, aiming to find specific states (ground state, excited state(s)), represent strongly correlated systems (typically found in metal-based compounds), or enhance computational efficiency through quantum embedding. In the set of single-reference post Hartree-Fock methods, coupled cluster is one of the main pillars in the quantum chemistry community. The wave function is written as an exponential ansatz,

\begin{equation}\label{eq:cc_cluster}
  \ket{\psi}=e^T|\psi_0\rangle = e^{T_1+T_2+T_3+T_4...} |\psi_0\rangle,
\end{equation}
where $|\psi_0\rangle$, and $T$ are respectively the Hartree-Fock state, and the cluster operator, i.e. a sum of single- ($T_1$), double- ($T_2$), and higher-order excitation operators ($T_{n\geq3}$). Practically, coupled cluster equations are solved after being truncated: for example, the Coupled Cluster Singles and Doubles (CCSD) cluster operator consider up to double-electron excitations, as in $T_1 + T_2$, where 

\begin{align}
  T_1 &= \sum_i \sum_a t_a^i \hat{a}^{\dagger}_a \hat{a}_i, \\
  T_2 &= \frac{1}{4} \sum_{i,j} \sum_{a,b} t_{ab}^{ij} \hat{a}^{\dagger}_a \hat{a}^{\dagger}_b \hat{a}_i \hat{a}_j.
\end{align} 

Here, $t$, $\hat{a}^{\dagger}$, $\hat{a}$ represent the amplitude, the fermionic creation operator, and the fermionic annihilation operator, respectively. The indices $i$ and $j$ ($a$ and $b$) refer to occupied (virtual) orbitals.

The operations available to a quantum computer include unitary gates and non unitary measurements. The unitary coupled cluster ansatzes, more precisely its trotterized version, was found to be a natural way of encoding coupled cluster wave functions in a variational manner. At the time of writing this article, the Unitary Coupled Cluster Singles and Doubles (UCCSD) ansatz has established itself as a standard in quantum chemistry calculations on quantum computers. As mentioned in the introduction, this workflow suffers from several flaws that hinders its use, as the VQE runtime would be prohibitive for quantum devices. Several research groups have worked on ways to mitigate those problems, including grouping Hamiltonian terms~\cite{yenDeterministic2023,bansinghFidelity2022}, removing Hamiltonian and/or excitation terms based on symmetries~\cite{bravyiTapering2017,setiaReducing2020,fischerSymmetry2019}, making hardware-efficient ansatzes~\cite{ryabinkinQubit2018,kandalaHardwareefficient2017}, constructing the circuit iteratively~\cite{grimsleyAdaptive2019,ryabinkinIterative2020}, and starting the optimization scheme with better initial parameters~\cite{McClean_2016,Romero_2019,hirsbrunner2023beyond}. Although much progress have been made, there are still obstacles in variational optimization, and an unfavorable scaling of quantum resources to obtain the ground state~\cite{leeEvaluating2023,mccleanTheory2016}.

The workflow proposed in this article maps coupled cluster amplitudes from Eq.~\ref{eq:cc_cluster} to a quantum circuit, making use of the existing efficient and parallelized code implementation of coupled cluster theory~\cite{osti_1817656}. \citeauthor{hirsbrunner2023beyond} have also explored a similar no-optimization strategy, using a sparse wave function circuit solver to map the CC amplitudes to UCC coefficients~\cite{hirsbrunner2023beyond}. In our strategy, amplitudes are normalized during the mapping procedure, as real-world probabilities must sum to one on the quantum device. The state preparation is then equivalent to the VCC approach, which computes the energy via a set of normalized coefficients. Similar to \citeauthor{hirsbrunner2023beyond}'s strategy, the correspondence between CCSD and our circuit preparation is expected to decrease when classical CCSD breaks down. This aspect of the work, and the differences between the two methods (UCC vs VCC)~\cite{harsha2018difference}, require further investigation beyond the scope of this article.

\subsection{Circuit primitives}\label{sec:circ_primitives}

Our goal is to show the construction of

\begin{equation}
    \ket{\psi} \rightarrow \frac{e^{\alpha \hat{a}^{\dagger}_p \hat{a}_q} \ket{\psi}}{\|e^{\alpha \hat{a}^{\dagger}_p \hat{a}_q} \ket{\psi} \|}.
\end{equation}

This mapping requires the implementation of a circuit applying

\begin{equation}\label{eq:exp_exc_op}
    e^{\alpha \hat{a}^{\dagger}_p \hat{a}_q} = \hat{I} + \alpha \hat{a}^{\dagger}_p \hat{a}_q.
\end{equation}

The creation $\hat{a}^{\dagger}_p$ and annihilation $\hat{a}_q$ operators for fermions can be mapped to Pauli matrices in the occupation basis via Eqs.~\ref{eq:adagtopauli}-\ref{eq:atopauli}.

\begin{align}
  \hat{a}^{\dagger}_p &= \left( \bigotimes_{j=1}^{p-1} Z_j \right) \otimes \frac{X_p-iY_p}{2}, \label{eq:adagtopauli}\\
  \hat{a}_q &= \left( \bigotimes_{j=1}^{q-1} Z_j \right) \otimes \frac{X_q+iY_q}{2}. \label{eq:atopauli}
\end{align}

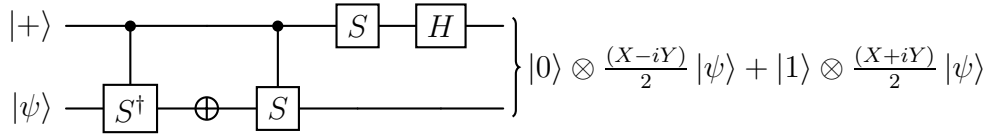
\begin{figure}
  \centering
  \begin{quantikz}
    \lstick{\ket{+}} & \ctrl{1} & & \ctrl{1} & \gate{S} & \gate{H} & \rstick[2]{$\ket{0} \otimes \frac{\left( X - i Y \right)}{2} \ket{\psi} + \ket{1} \otimes \frac{\left( X + i Y\right)}{2} \ket{\psi}$} \\
    \lstick{\ket{\psi}} & \gate{S^{\dagger}} & \targ{} & \gate{S} & & &
  \end{quantikz}
  \caption{Simplest circuit primitive for a fermionic operator. If the \ket{0} (\ket{1}) state is measured on the first qubit, a creation (annihilation) operator is applied on \ket{\psi}.}
  \label{fig:simplest_primitive}
\end{figure}

As shown in Appendix~\ref{app:fermop_using_ancillas}, the circuit in Fig.~\ref{fig:simplest_primitive} is capable of applying fermionic operators in a non-deterministic way. Therefore, one can choose which operator to apply, depending on the measurement outcome of the ancilla qubit. This can be ensured via post-processing on the \ket{0} or \ket{1}, or by only post-processing on \ket{0} and adapting the last phase gate on the ancilla qubit to be either $S$ or $S^{\dagger}$.

This simple primitive can be extended to an n-electron excitation in a trivial way by doing the same operation in parallel on the relevant qubits. Now that we have a way to map an arbitrary n-electron excitation term onto a quantum circuit, we must take care of the amplitude injection, as required by Eq.~\ref{eq:exp_exc_op}. This is accomplished by adding a control qubit initiated in an arbitrary state $\cos{\theta}\ket{0} + e^{i\phi}\sin{\theta} \ket{1}$. After measurement of this qubit in the $\{ \cos{\theta}\ket{0} + \sin{\theta}\ket{1}, -\sin{\theta}\ket{0} + \cos{\theta}\ket{1}\}$ basis, selecting the \ket{0} projects the state to

\begin{equation}\label{eq:cc_amp_to_param}
    \frac{\cos^2{(\theta/2)} \ket{\psi} \pm \sin^2{(\theta/2)} \hat{a}_{p}^{\dagger} \hat{a}_q \ket{\psi}}{\sqrt{\cos^4{(\theta/2)} + \sin^4{(\theta/2)}}} \approx \left(I + \alpha \hat{a}_{p}^{\dagger} \hat{a}_q\right) \ket{\psi}.
\end{equation}

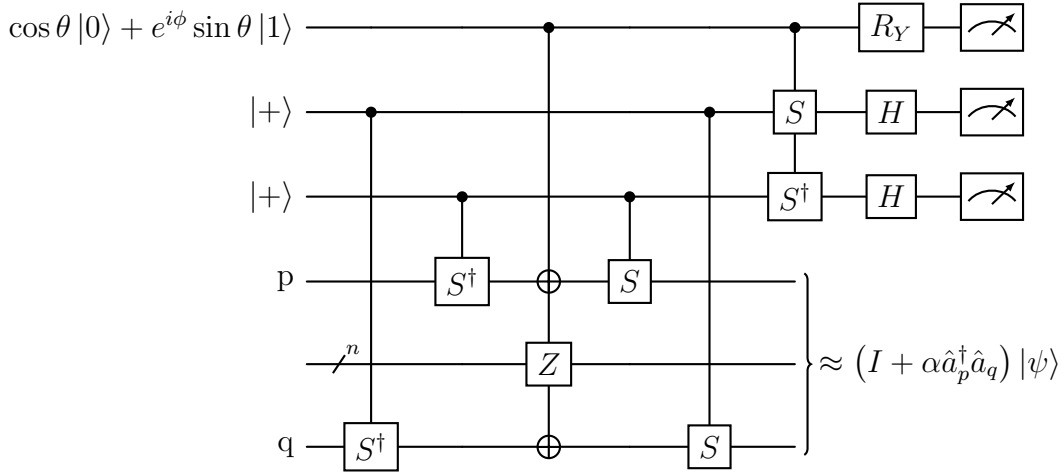
\begin{figure}
  \centering
  \begin{quantikz}
    \lstick{$\cos{\theta}\ket{0} + e^{i\phi}\sin{\theta} \ket{1}$} & & & \ctrl{5} & & & \ctrl{2} & \gate{R_Y} & \meter{} \\
    \lstick{\ket{+}} & \ctrl{4} & & & & \ctrl{4} & \gate{S} & \gate{H} & \meter{} \\
    \lstick{\ket{+}} & & \ctrl{1} & & \ctrl{1} & & \gate{S^{\dagger}} & \gate{H} & \meter{} \\
    \lstick{p} & & \gate{S^{\dagger}} & \targ{} & \gate{S} & & \rstick[4]{$\approx \left(I + \alpha \hat{a}_{p}^{\dagger}\hat{a}_q\right) \ket{\psi}$} \\
    &\qwbundle{n} & & \gate{Z} & & & \\
    \lstick{q} & \gate{S^{\dagger}} & & \targ{} & & \gate{S} & \\
  \end{quantikz}
  \caption{Circuit implementation of $e^{\alpha \hat{a}^{\dagger}_p \hat{a}_q}$. The ancilla qubit has the purpose of injecting the operator amplitude $\alpha$. The second and third ancilla qubits are used for the fermionic excitation operator implementation. The other qubits encode the molecular wave function.}
  \label{fig:single_exc_circuit}
\end{figure}

This is the core pattern (Fig.~\ref{fig:single_exc_circuit}) for the construction of fermionic excitations in the occupation basis. Generalization of this construction can be made to perform n-electron excitation, with the use of $2n +1$ ancilla qubits. If $\alpha$ is small, the generated state should be close to the classical CCSD state.

\subsection{Full circuit construction}\label{sec:full_cc_circuit}

\begin{figure}
  \centering
  \begin{quantikz}
    \lstick{\ket{0}} & \gate{U_1} & \ctrl{1} & \gate{R_{Y1}} & \meter{} & \wireoverride{n} & \lstick{\ket{0}}\wireoverride{n} & \gate{U_2}  & \ctrl{1} & \gate{R_{Y2}}  & \meter{} \\
    \lstick{\ket{+}} & \qwbundle{2} & \gate[2]{\hat{a}^{\dagger}_p \hat{a}_q} & \gate{H} & \meter{} & \wireoverride{n} & \lstick{\ket{+}}\wireoverride{n} & \qwbundle{4} & \gate[2]{\hat{a}^{\dagger}_p \hat{a}_q \hat{a}^{\dagger}_r \hat{a}_s} & \gate{H}  & \meter{} \\
    \lstick{\ket{\psi_0}} &  & & & & & & & & & & \rstick{\ket{\psi}}
  \end{quantikz}
  \caption{Example of a full quantum circuit construction with a single and double-electron excitation terms. Single-electron terms require 3 ancilla qubits, while double-electron terms 5 ancilla qubits. $U_i$ can be decomposed into a sequence of $R_Y$ and phase gates to prepare the ancilla qubit in the expected state. The $R_{Yi}$ gate are used to transform the measurement basis. After each mid-circuit measurement, those ancilla qubits can be reused if the reset instruction is fast enough.}
  \label{fig:full_cc_circuit}
\end{figure}
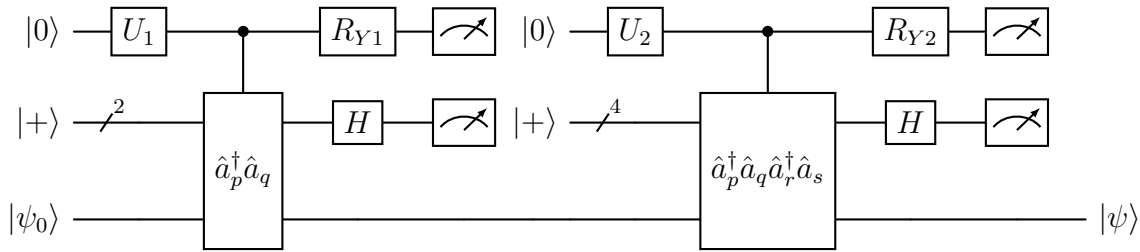

From the building blocks established in the previous section, we envision the construction of the full CC circuit to be as illustrated in Fig.~\ref{fig:full_cc_circuit}.

$U_i$ prepares the $\cos{\theta}\ket{0} \pm \sin{\theta} \ket{1}$ state that serves to inject the $\alpha$ coefficient, and it can be decomposed into a sequence of $R_Y$ and phase gates. The $R_{Yi}$ gate makes the measurement in the appropriate basis mentioned in Section~\ref{sec:circ_primitives}. Each coupled cluster operator is applied sequentially, and the mid-circuit measurement outcome indicates whether the operation was successful or not.

\section{Evaluation of Non-unitary Coupled Cluster Circuits}\label{sec:results}

This section presents the results to assess the correctness and scalability of our non-unitary coupled cluster circuits. Our aim is to demonstrate that our method correctly maps the classical state into a quantum circuit using mid-circuit measurements. To measure this, we compare the energy evaluation from the circuit to the corresponding classically calculated CC energy for a set of small systems. In the remainder of this section, we provide the computational cost of this method compared to an equivalent VQE-UCCSD workflow.

\subsection{CCSD amplitudes mapping}

Computation of the CCSD solutions were performed through the PySCF code~\cite{sunRecent2020}. The molecular dataset consists of simple molecules, ranging from hydrogen chains to few-atoms molecules. The atomic coordinates were retrieved from the CCCBDB database~\cite{cccbdb_database}, and the STO-3G minimal basis set was used for all of the use cases. The Jordan-Wigner qubit mapping was used~\cite{jordanUber1928,nielsen2005fermionic}, resulting in systems ranging from 4 to 16 qubits. Accounting for the additional ancilla qubits for the mid-circuit measurement, the circuit sizes extend from 9 to 21 qubits. CCSD calculations, fermionic to qubit transformations, circuit construction, and energy evaluations were all done using the Tangelo package~\cite{tangelo}, with Qulacs as a quantum circuit simulation backend~\cite{suzuki2021qulacs}. 

\begin{figure}
  \centering
  \includegraphics{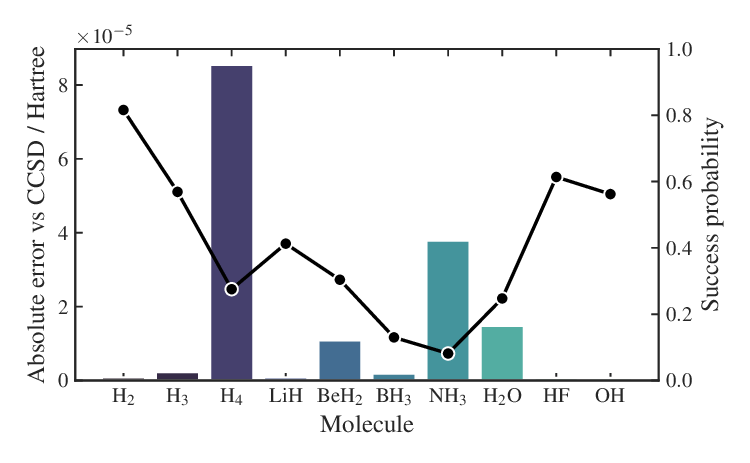}
  \caption{Set of molecular systems (in STO-3G) for the conversion of CCSD amplitudes into quantum circuits. The bar plot corresponds to a single circuit energy evaluation (left axis). The scatter plot refers to the success probability (right axis).}
  \label{fig:benchmark_molecules}
\end{figure}

After the conversion of the amplitudes to rotational parameters with Eq.~\ref{eq:cc_amp_to_param}, the circuits were built following the protocol described in Sections~\ref{sec:circ_primitives} and~\ref{sec:full_cc_circuit}. The resulting energy evaluation errors (versus classical CCSD) of the encoded wave functions are displayed in Fig.~\ref{fig:benchmark_molecules}.

The energy difference between the classical CCSD calculations and the circuit Hamiltonian evaluations are within $(1.6 \pm 2.6)\times 10^{-5}$~Ha for this dataset. The minimum error observed is $7.0\times10^{-7}$~Ha, and the maximum is $8.5\times10^{-5}$~Ha. While those results aren't numerically exact, the authors argue that they are within acceptable boundaries, especially considering the target chemical accuracy is generally 1.6 mHa~\cite{pople1999nobel}. The discrepancy encountered can be attributed to the approximation raised in Eq.~\ref{eq:cc_amp_to_param}. In our circuit construction, the coefficient in front of the identity operator isn't 1 due to the normalisation with $\alpha$. If $\alpha$ is small, the amplitude mapping should be close to the non-variational coupled cluster state.

The energy evaluation for the coupled cluster state was conducted for a successful measurement run. This means that each individual circuit primitive successfully outputs the $\ket{0}$ state after measurement. If this is the case, it means that all fermionic excitation terms have been applied onto the state correctly. This probability is dependent on the amplitude magnitude $\alpha$ and the current state symmetry, for example if applying $\hat{a}^{\dagger}_q a_p$ (corresponding to a $\ket{11}$ measurement on the $\ket{+}$-initialized qubits) is giving a valid fermionic state. Ensuring that the $\ket{+}$-initialized qubits all give the $\ket{0}$ state means that the chemical symmetries are kept consistent with the Hartree-Fock state. The failure probability of applying the expected operation is $\mathcal{O}(\sin^2{(\theta)}\cos^2{(\theta)}) = \mathcal{O}(\sin^2{(2\theta)})$, which is equivalent to $\mathcal{O}(|\alpha|^2)$ for small $\alpha$. This probability also depends on the reference state used. In our case, the Hartree-Fock state is used, and our numerical results show success probabilities ranging from 8\% to 80\%. Therefore, there is a significant chance of producing the wanted state with a reasonable amount of device shots.

The success probability could be enhanced via multiple strategies with different complexities. For example, simply removing low-amplitude excitation terms should eliminate failing points~\cite{mehendale2023exploring}, and using techniques like resource-saving compilation algorithm~\cite{chen2021concentration} or performing amplitude amplification on the ancilla qubits~\cite{brassard1997exact,grover1998quantum,berry2014exponential} could boost the success probability. The latter has been already explored for probabilistic imaginary-time evolution~\cite{nishi2023quadratic}.

\subsection{Scaling discussion}\label{sec:scaling}

Efficient initial state preparation is crucial to reduce QPE runtime for chemical Hamiltonians~\cite{fomichevInitial2023}. For a given overlap with the true ground state, there are benefits to reducing the overhead of initial state preparation, mainly for minimizing the time-to-solution. In this section, we show the scaling cost for initial state preparation when using non-unitary circuits and compare it to the unitary version in the context of a CC wave function. Scaling will be discussed from three angles: CNOT gate ratios, T gate ratios, and circuit representation of a converged coupled cluster state.

\begin{figure}
  \centering
  \includegraphics{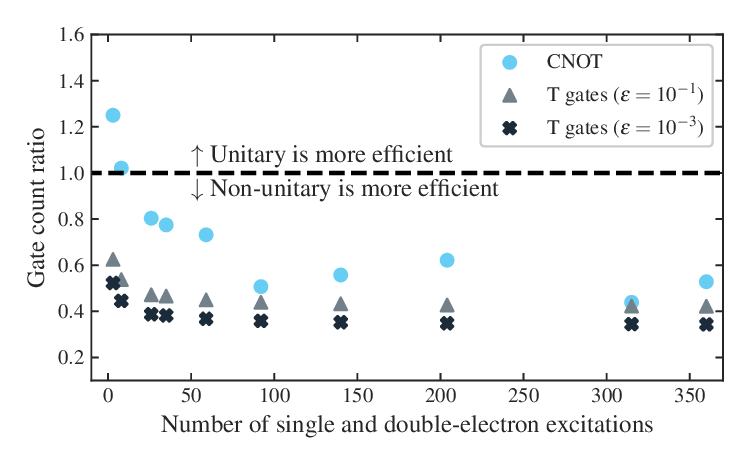}
  \caption{Gate count ratios for the same system reported in Fig.~\ref{fig:benchmark_molecules}. The non-unitary strategy requires fewer gates when the ratio is below unity. The data used for computing plotting this figure is shown in Table~\ref{tab:data_scaling_gates}.}
  \label{fig:scaling_gates}
\end{figure}

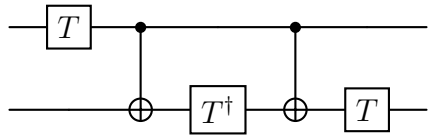
\begin{figure}
  \centering
  \begin{quantikz}
     & \gate{T} & \ctrl{1} & & \ctrl{1} & & \\
     & & \targ{} & \gate{T^{\dagger}} & \targ{} & \gate{T} &
  \end{quantikz}
  \caption{Equivalent circuit using two CNOTs for a controlled S gate.}
  \label{fig:cs_to_cnots}
\end{figure}

Quantum scaling can be evaluated from the perspective of the CNOT count, as this is a significant bottleneck in the implementation of circuits representing strongly-correlated fermionic systems. Fig.~\ref{fig:scaling_gates} is showing the CNOT ratios (non-unitary / unitary) for the molecular systems previously used as the test set in Fig.~\ref{fig:benchmark_molecules}. The x-axis of the plot displays the total number of single and double excitations for each molecule. Those were computed from the resource estimation when considering all single and double-electron excitation terms, which is a worst case scenario since any gates associated with small parameter excitations can effectively be removed from the circuit. For the trotterized UCCSD, the conventional construction recipe with CNOT ladders was used~\cite{whitfieldSimulation2011}. For the non-unitary circuit construction, circuits from Figs~\ref{fig:single_exc_circuit} and~\ref{fig:full_cc_circuit} were used, and two CNOTs were counted for each controlled S gate using the circuit in Fig.~\ref{fig:cs_to_cnots}. We also show the T gate ratios, as it is a relevant metric in the context of state preparation for fault-tolerant quantum computing. For the non-unitary circuits, T gates are required for the controlled-S operations, and for the two rotation gates used for injecting the amplitudes. Using results from Ref.~\citenum{kliuchnikov2023shorter}, the amount of T gates can be approximated by $1.12 \log_2{(1/\epsilon)} + 18n + 10.6$ for an $n$-body excitation. For the UCC variant, an $n$-body excitation requires $2^{2n-1}$ rotations, which results in a T gate count of about $2^{2n-1}(5.3+0.56\log_2{(1/\epsilon)})$. The ratios of those T gate counts are plotted in Fig.~\ref{fig:scaling_gates} for several $\epsilon$.

Across the set, the non-unitary strategy tends to be more efficient in terms of CNOT gate usage compared to the unitary strategy by showing an average reduction of 28\%. This reduction goes up by 41\% when removing the molecular system containing only hydrogen atoms. The T gate ratios are also in favor of the non-unitary state preparation protocol, which show a reduction of 53\% and 61\% (57\% on average) for $\epsilon=10^{-1}$ and $\epsilon=10^{-3}$, respectively. The gain is made by circumventing the need of implementing $2^{2n-1}$ rotation gates for $n$-body excitations, which is particularly prohibitive for higher-order terms. The introduction of triple excitation terms might lead to chemical accuracy~\cite{gonthierMeasurements2022}, and several groups have taken a shot at implementing ans\"{a}tze with triple excitation terms~\cite{haidarExtension2023,windomNew2024}. While also considering the existence of massively parallel implementation of the CCSD(T) code~\cite{osti_1817656}, it would be worthwhile to pursue an investigation to consider higher-order terms with our circuit design.

\begin{table}
  \centering
  \caption{Gate counts for the non-unitary and VQE-UCCSD final circuits}\label{tab:data_scaling_gates}
  \begin{tabular}{l|cc|ccc|ccc}
    & & & \multicolumn{3}{c|}{Non-unitary} & \multicolumn{3}{c}{VQE-UCCSD} \\
    Molecule & $N_{\text{singles}}$ & $N_{\text{doubles}}$ & CNOTS & T gates & T gates & CNOTS & T gates & T gates \\
    & & & & $\epsilon=0.1$ & $\epsilon=0.001$ & & $\epsilon=0.1$ & $\epsilon=0.001$ \\ \hline
    \ce{H2} & 2 & 1 & 60 & 120 & 138 & 48 & 192 & 264 \\
    \ce{H3} & 4 & 4 & 98 & 344 & 392 & 96 & 640 & 880 \\
    \ce{H4} & 8 & 18 & 540 & 1208 & 1364 & 672 & 2560 & 3520 \\
    \ce{LiH} & 16 & 76 & 1444 & 4496 & 5048 & 2848 & 10240 & 14080 \\
    \ce{BeH2} & 24 & 180 & 1830 & 10176 & 11400 & 2944 & 23808 & 32736 \\
    \ce{BH3} & 32 & 328 & 4734 & 18144 & 20304 & 8960 & 43008 & 59136 \\
    \ce{NH3} & 30 & 285 & 6162 & 15840 & 17730 & 14032 & 37440 & 51480 \\
    \ce{H2O} & 20 & 120 & 1820 & 6920 & 7760 & 3264 & 16000 & 22000 \\
    \ce{HF} & 10 & 25 & 632 & 1640 & 1850 & 816 & 3520 & 4840 \\
    \ce{OH} & 13 & 46 & 644 & 2834 & 3188 & 880 & 6304 & 8668
  \end{tabular}
\end{table}

The state preparation protocol of a VQE workflow would typically involve:
\begin{enumerate}
    \item\label{enu:first_vqe_step} The classical computation of an initial set of variational parameters, which are typically derived from a low-cost quantum chemistry method (Hartree-Fock, M{\o}ller–Plesset perturbation theory, etc.).
    \item\label{enu:energy_eval} Performing an energy evaluation, which requires measurement of a set of single-qubit Pauli bases.
    \item Repeating step~\ref{enu:energy_eval} many times to numerically compute an energy gradient (or whatever quantity that helps decide how each numerical parameter should be updated).
    \item\label{enu:last_vqe_step} Updating the variational parameters.
    \item Finally, performing steps~\ref{enu:first_vqe_step} to~\ref{enu:last_vqe_step} until convergence is reached according to the predefined criteria.
\end{enumerate} 

After the protocol, the UCCSD ground state is obtained deterministically, at the cost of performing many energy evaluations. Considering the importance of initial parameterization and the problem of Barren plateaus, the search of the parametric circuit corresponding to a global minimum is not guaranteed to converge, as the eigenvalue landscape is a non-convex function in the multi-dimensional Hilbert space.

In comparison, our state preparation protocol using mid-circuit measurements implies:
\begin{enumerate}
    \item The computation of coupled cluster amplitudes with a classical quantum chemistry package.
    \item The conversion of those amplitudes into rotation angles with Eq.~\ref{eq:cc_amp_to_param}.
    \item And the execution of the circuit as shown in Fig.~\ref{fig:full_cc_circuit} until all measures output 0. If it fails, the circuit is executed again from the beginning. 
\end{enumerate}

One advantage of the non-unitary protocol is that the circuit is pre-parameterized through an efficient existing classical algorithm. Most popular classical quantum chemistry software include an implementation of coupled cluster singles and doubles, from which the coupled cluster amplitudes are computed efficiently with a scaling of $\mathcal{O}(n^6)$, where $n$ is the number of spin orbitals in the molecule. As a result, it removes the need to perform energy estimates to construct a high quality state, where the number of measurements of each estimate scale as $\mathcal{O}(n^3)$ by employing state-of-the-art tomography methods on the quantum computer~\cite{yen2023deterministic}. The trade-off is having a state protocol which is probabilistic.

For the previously introduced molecular test set, we argue that the failure probability is low enough that repeating the state preparation up until success is a viable option, especially when factoring the energy evaluation overhead cost that is being avoided. This is achieved while still producing quantum states, in comparison to the VQE-UCCSD protocol, having similar overlaps with their corresponding ground state, as shown in Table~\ref{tab:overlaps}.

\begin{table}
  \centering
  \caption{Overlaps, versus the ground state, of the relevant wave functions discussed in this article}\label{tab:overlaps}
  \begin{tabular}{l|ccc}
    Molecule & Hartree-Fock & Non-unitary & VQE-UCCSD \\ \hline
    \ce{H2} & 0.993615 & 1.000000 & 1.000000 \\
    \ce{H3} & 0.971278 & 0.987103 & 0.987104 \\
    \ce{H4} & 0.967711 & 0.999972 & 0.999979 \\
    \ce{LiH} & 0.987154 & 1.000000 & 1.000000 \\
    \ce{BeH2}& 0.986104 & 0.999842 & 0.999845 \\
    \ce{BH3} & 0.990903 & 0.999955 & 0.999956 \\
    \ce{NH3} & 0.991116 & 0.999991 & 0.999991 \\
    \ce{H2O} & 0.986684 & 0.999980 & 0.999980 \\
    \ce{HF} & 0.992431 & 1.000000 & 0.999998 \\
    \ce{OH} & 0.992229 & 1.000000 & 0.999998 \\
  \end{tabular}
\end{table}

\section{Conclusion}

We proposed a new circuit construction protocol to map coupled cluster equations and amplitudes to gate-based quantum devices. This protocol is probabilistic, however the probabilities are within the realms of values where observing successful state preparation is achieved. When it succeeds, the output state should give a similar energy to the classically optimized coupled cluster amplitudes. In comparison with the usual VQE-UCCSD protocol, it avoids the search of optimized parameters for the parameterized ansatz, which would require long coherence time and rapid connection between the classical and quantum apparatus. This work serves as a starting point for further quantum treatment of a correlated wave function, like improving the wave function accuracy by considering higher-order excitation terms~\cite{haidarExtension2023,windomNew2024}, or probing excited states~\cite{stanton1993equation,ollitrault2020quantum}.

\appendix

\section{Appendix}

\subsection{Implementation of a fermionic operator using an ancilla qubit}\label{app:fermop_using_ancillas}

Here is the step-by-step reasoning of the circuit implementation as shown in Fig.~\ref{fig:simplest_primitive}.

\begin{align*}
  \ket{+} \otimes \ket{\psi} &= \frac{1}{\sqrt{2}} \left( \ket{0} + \ket{1} \right) \otimes \ket{\psi} \\
  &\xrightarrow{CS^{\dagger}} \frac{1}{\sqrt{2}} \left( \ket{0} \otimes \ket{\psi} + \ket{1} \otimes S^{\dagger}\ket{\psi} \right)  \\
  &\xrightarrow{X} \frac{1}{\sqrt{2}} \left( \ket{0} \otimes X\ket{\psi} + \ket{1} \otimes XS^{\dagger}\ket{\psi} \right) \\
  &\xrightarrow{CS} \frac{1}{\sqrt{2}} \left( \ket{0} \otimes X\ket{\psi} + \ket{1} \otimes SXS^{\dagger}\ket{\psi} \right) \\
  &\xrightarrow{S^{\dagger}} \frac{1}{\sqrt{2}} \left( S^{\dagger}\ket{0} \otimes X\ket{\psi} + S^{\dagger}\ket{1} \otimes SXS^{\dagger}\ket{\psi} \right) = \frac{1}{\sqrt{2}} \left( \ket{0} \otimes X\ket{\psi} - i\ket{1} \otimes SXS^{\dagger}\ket{\psi} \right) \\
  &\xrightarrow{H} \frac{1}{\sqrt{2}} \left( \frac{1}{\sqrt{2}} \left( \ket{0} + \ket{1} \right) \otimes X\ket{\psi} \mp \frac{i}{\sqrt{2}} \left( \ket{0} - \ket{1} \right) \otimes SXS^{\dagger}\ket{\psi} \right) \\
  &= \frac{1}{2} \left( \ket{0} \otimes \left( X - i SXS^{\dagger} \right) \ket{\psi} + \ket{1} \otimes \left( X + i SXS^{\dagger}\right) \ket{\psi} \right) \\
  &= \ket{0} \otimes \frac{\left( X - i Y \right)}{2} \ket{\psi} + \ket{1} \otimes \frac{\left( X + i Y\right)}{2} \ket{\psi},
\end{align*}

Where we used the identity $SXS^{\dagger} = Y$.

\bibliography{nuccsd.bib}

\end{document}